\pdfoutput=1
\documentclass[ERTS, 2026]{ERTS} 

\usepackage{graphicx}
\usepackage{amsmath}
\usepackage[utf8]{inputenc}  
\usepackage[english]{babel}                             
\usepackage{syntax}
\usepackage{tikz}
\usepackage{soul,hyperref}
\usepackage{xcolor,xspace}
\usepackage{listings}
\usepackage{amsfonts}
\usepackage{amsmath}
\usepackage[font=small,skip=1pt]{caption}
\usepackage{graphicx} 

\usepackage{subcaption} 
\usepackage{pifont}

\usetikzlibrary{shapes.arrows}

\newtheorem{definition}{\textbf{Definition}}
\newtheorem{property}{\textbf{Property}}
\newtheorem{example}{\textbf{Example}}
\newtheorem{remark}{\textbf{Remark}}
\newtheorem{assumption}{\textbf{Assumption}}

\newcommand{\sizeOnChip}{\emph{size$_{MEM}$}\xspace}
\newcommand{\nbmac}{\emph{nbop$_{PE}$}\xspace}
\newcommand{\actions}{\emph{Actions}}

\newcommand{\durationstep}{\emph{$\delta$}}

\newcommand{\setPixelPatch}{\emph{pxl\_in\_P}}
\newcommand{\PatchToGroup}{\emph{P\_g}}
\newcommand{\PixelToGroup}{\emph{pxl\_g}}
\newcommand{\PixelOverlapped}{\emph{pxl\_ovlp}}

\newcommand{\PixelIslice}{\emph{pxl\_I}}
\newcommand{\NbPixelOverlapped}{\emph{nb\_pxl\_ovlp}}

\makeatletter

\AddToHook{cmd/maketitle/before}{\let\thetitle\@title}
\makeatother

\begin{document}

\title{Convolutions Predictable Offloading  to an Accelerator: Formalization and Optimization}

\author{%
	Benjamin Husson\authorNumber{1}, Mohammed Belcaid\authorNumber{2}, Thomas Carle\authorNumber{3}, and Claire Pagetti\authorNumber{4}
}

\address{
	\affiliation{{1,2}}{CS Group, Toulouse, France}
	\affiliation{3}{Université de Toulouse -- IRIT, Toulouse, France}
   	\affiliation{4}{ONERA, Toulouse, France}
}

\maketitle

\chead{\thetitle}

\pagestyle{fancy}

\thispagestyle{plain}

\licenseFootnote{Benjamin Husson et al}





\maketitle

\begin{abstract}%
Convolutional neural networks (CNNs) require a large number of multiply-accumulate (MAC) operations. To meet real-time constraints, they often need to be executed on specialized accelerators composed of an on-chip memory and a processing unit. However, the on-chip memory is often insufficient to store all the data required to compute a CNN layer. Thus, the computation must be performed in several offloading steps. We formalise 
such sequences of steps and apply our formalism to a state of the art decomposition of convolutions. In order to find optimal strategies in terms of duration, we encode the problem with a set of constraints. A Python-based simulator allows to analyse in-depth computed strategies. 
\end{abstract}


\section{Introduction}

\begin{figure}[hbt]
    \centering
    \resizebox{.75\linewidth}{!}{\begin{tikzpicture}[node distance=1.5cm, very thick]

     \tikzstyle{txt} = [style={text width=6cm,align=left}]
     \tikzstyle{block} = [draw,minimum height=4em,
     minimum width=4em, inner sep=5pt];

     \draw (0,0) node[block] (cpu){CPU};
     \path (cpu)+(1.25,0)
      node[double arrow, draw=black,  minimum width = 10pt, single arrow head extend=1mm,
      minimum height=10mm] {}; 
    \path (cpu)+(2.5,0) node[block] (dram){DRAM};
    \path (dram)+(1.25,0) 
     node[double arrow, draw=black,  minimum width = 10pt, single arrow head extend=1mm,
      minimum height=10mm] {}; 
    \path (dram)+(2.7,0) node[block] (onchipm){MEM};
    \path (onchipm)+(2.5,0) node[block] (calc){PE};
    \path (onchipm)+(1.25,0)  node[double arrow, draw=black,  minimum width = 10pt, single arrow head extend=1mm,
      minimum height=10mm] {}; 
    \path (dram)+(3.9,0) node[block,minimum height=6.5em,
     minimum width=12.5em] (rect){};

\path (onchipm)+(0,1) node {\textbf{Accelerator}};
\end{tikzpicture}
     }
    \caption{Generic accelerator architecture}
    \label{fig:archi-paper}
\end{figure}
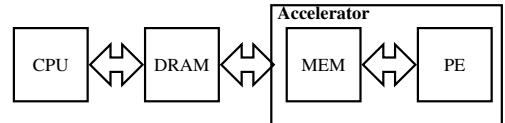

The advent of Machine Learning has opened unprecedented possibilities across the industry.
Among Machine Learning models, Convolutional Neural Networks (CNNs) have proven their capabilities in image segmentation, classification, and detection.
This work focuses on the use of off-line trained CNNs for 
safety-critical real-time applications.
A first observation is that 
convolutions and dense layers
often result in a large number of 
MAC (multiply-accumulate) operations,
leading to high computational demand. 
Thus, to achieve reasonable temporal performance, 
CNNs often need to be executed on specialized accelerators such as Graphics Processing Units \cite{GPU-CNN}, 
Tensor Processing Units (TPUs) \cite{TPU}, Neural Processing Units \cite{npu-example}, and FPGA-based accelerators \cite{VTA-paper} \cite{HP-FPGA} \cite{li2025designimplementationfpgabasedhardware}.
Such accelerators may, for instance, contain 
parallel MAC hardware units 
that permit efficient MAC operations execution.
Figure \ref{fig:archi-paper}
illustrates a generic accelerator architecture where
\emph{MEM} is the on-chip memory and \emph{PE} is the processing part.
Data are exchanged between the CPU and the accelerator via an off-chip DRAM.

\subsection{Problem statement} 
\begin{figure}[h!bt]
\centering
    
    \begin{subfigure}[b]{0.35\textwidth}
        \centering
        \includegraphics[width=\textwidth]{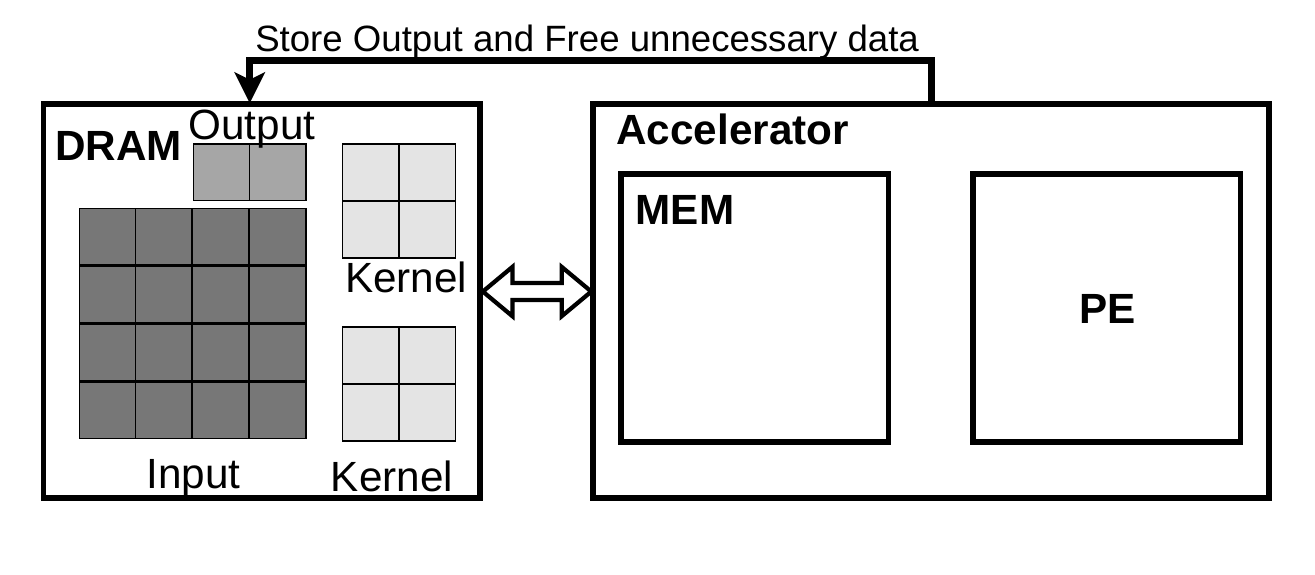} %
        \caption{1) free the memory and 2) store the results}
        \label{fig:seq5}
    \end{subfigure}

    \begin{subfigure}[b]{0.35\textwidth}
        \centering
        \includegraphics[width=\textwidth]{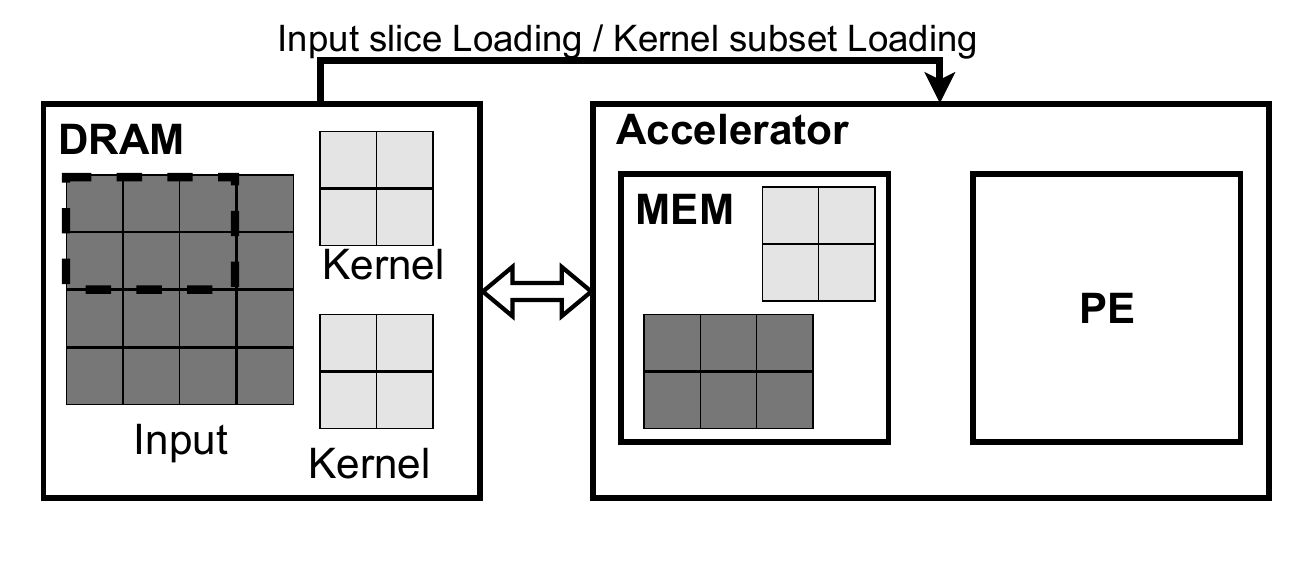} %
        \caption{3) load the elements}
        \label{fig:seq2}
    \end{subfigure}

    \begin{subfigure}[b]{0.35\textwidth}
        \centering
        \includegraphics[width=\textwidth]{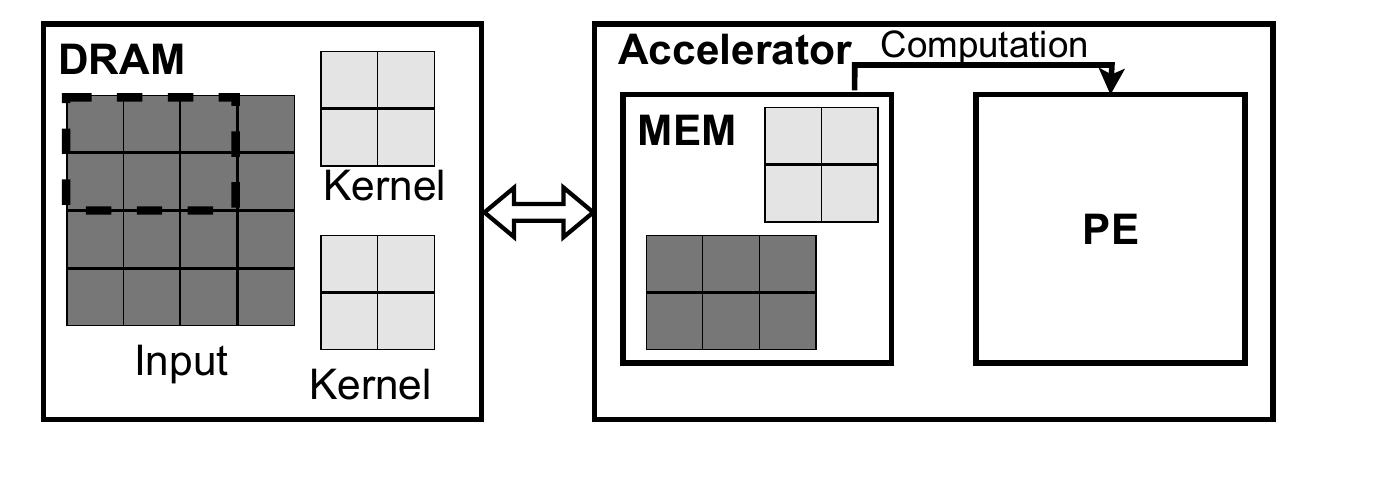} %
        \caption{4) launch the computation}
        \label{fig:seq3}
    \end{subfigure}
    \caption{A step = sequence of execution}
    \label{fig:allsequences}  
\end{figure}

Usually, the on-chip memory is rather small, preventing complete local storage of both the CNN parameters and the input needed to realize the computation. Consequently, the computation 
has to be done in several steps  \cite{Slice-kernel}.
Here, a step corresponds to a sequence of:
1) invalidating / freeing the memory from the loaded elements;
2) optionally writing back the computed value to the DRAM;
3) loading a slice of the input 
and a subset of the parameters (e.g., kernels for a convolution or weights for a dense layer);
4) performing the partial computation 
involving the slice and the subset of parameters;
5) and looping back to 1) as long as needed.
Figure \ref{fig:allsequences} illustrates the sequence involved in a step.

The CNNs we consider contribute to safety-critical real-time applications (e.g. drone detection \cite{belcaid-erts} or visual-based landing \cite{ducoffe-lard}).
 Consequently, 
 it is mandatory to estimate the time required to execute the model, which amounts to estimating the latency of each step and their combination.
To do so, we rely on a formal model of the offloading process.

\subsection{Contributions} 
In this work, we address the offloading  of convolutional layers to specialized accelerators when the on-chip memory is insufficient to contain a layer, meaning that the computation must be performed in multiple steps.
Our contributions include:
\begin{itemize}
    \item 
    The formalization of the notion of \emph{strategy}, where a strategy is a sequence of steps. This requires defining how the input and the kernels are decomposed per step, 
    quantifying the induced memory transactions between on-chip memory and DRAM, 
    and tracking the on-chip memory footprint.

    \item 
    The application of our formalism to one of the four strategies proposed in \cite{siu-memory-2018}, resulting in a formal strategy denoted further S1-baseline.
    Contrary to our work, which considers a fixed accelerator and searches for end-to-end durations, the purpose of the authors was to minimize the on-chip memory size and reduce the bandwidth between the DRAM and the on-chip memory. 

    \item 
    The translation of S1 into an ILP problem in charge of finding an optimal strategy (in terms of duration) for a given accelerator and a given convolutional layer to be mapped.

    \item
    A Python-based simulator to track which data are loaded, freed, computed, and what is stored in the on-chip memory at each step for a user defined strategy.
\end{itemize}

\subsection{Applicability of the proposed approach}
The purpose of the approach is manifold.
Even though the formalism remains at a high level of abstraction, it can model 
state-of-the-art execution strategies and hardware architectures. 
This, for instance, enables the comparison between different solutions with respect to the end-to-end duration criteria. The simulator complements the analysis by offering visual inspection of the behaviours (e.g. showing which data are loaded, freed, computed and written at each step), and a detailed description of the internal state of the accelerator (memory footprint of the input, kernels and output at each step).
The ILP formulation and two heuristic-based strategies offer an efficient way to help designers deploy convolution layers.

\begin{figure}[hbt]
    \centering
     \includegraphics[width=0.7\linewidth]{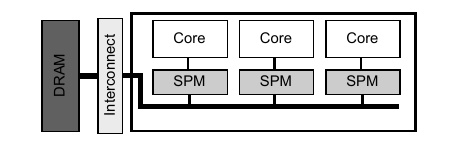}
    \caption{Multi-core with local SPM (e.g. AURIX)}
    \label{fig:aurix}
\end{figure}

We detail several hardware architectures that comply with our work.
Daini et al. \cite{DainiLZH25}
deploy CNNs on multi-core processors which come with local scratch-pad memory (SPM) and a shared DRAM as illustrated in Figure \ref{fig:aurix}. 
We can map this architecture to our generic one of Figure \ref{fig:archi-paper}: 
the on-chip memory corresponds to
the union of the local SPMs 
and the processing part is the union of the cores.
Contrary to us, the authors of \cite{DainiLZH25} consider that a step is the execution of a full layer (e.g. a convolution). We can thus complete their work by decomposing layers themselves into a succession of steps. In that case, 
both our notion of strategy and our ILP problem can help find efficient strategies.

\begin{figure}[hbt]
    \centering
    \includegraphics[width=0.6\linewidth]{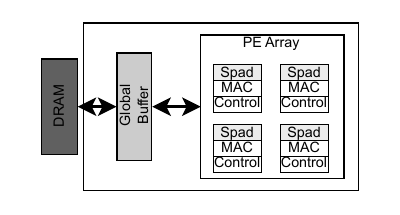}
    \caption{Eyeriss architecture}
    \label{fig:eyeriss}
\end{figure}

Chen et al. \cite{Eyeriss} introduced and implemented the Eyeriss accelerator, see Figure \ref{fig:eyeriss}, with the objective of optimising the dataflow for convolutional neural networks. Our formalism can be applied at different levels of the memory hierarchy:
by considering the on-chip memory to be either the global buffer or the 
union of the scratch-pad memories. The processing part is the PE array.
In addition, the hardware architecture enables the computation of 2D convolution by dividing it into 1D convolution primitives.
Thus, the ILP problem is also applicable. 

\begin{figure}[hbt]
    \centering
    \includegraphics[width=0.6\linewidth]{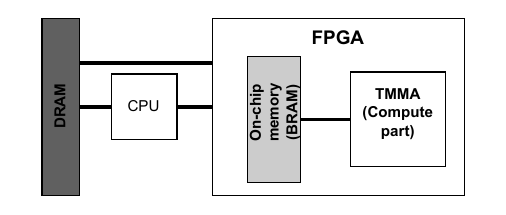}
    \caption{TMMA architecture}
    \label{fig:TMMA}
\end{figure}

Another example is the FPGA-based accelerator, named TMMA (see Figure \ref{fig:TMMA}), of \cite{li2025designimplementationfpgabasedhardware} 
featuring matrix multiplications. It means that convolutions are translated into matrix multiplication operations and more precisely block GeMM (General Matrix Multiplication) algorithm. 
Practically,  to perform $C=A\times B+C$, the matrices are sliced into tiles and the intermediate results are accumulated. It means that such operation is performed in several computation steps. 
Our formalism can be applied to this architecture  by considering the BRAM as the on-chip memory, the DRAM as the off-chip memory and the computation as a matrix multiplication (we do not model the sub-decomposition into tiles). 
However, we cannot use our detailed strategies (heuristic or those found with the ILP solver) because convolutions are translated into matrix operations. We need to slightly adapt our ILP problem.

\begin{figure}[hbt]
    \centering
    \includegraphics[width=0.6\linewidth]{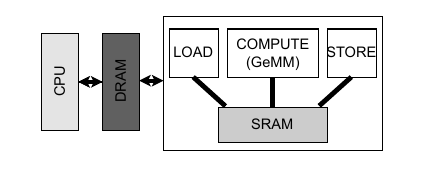}
    \caption{VTA architecture}
    \label{fig:VTA}
\end{figure}

The VTA (Versatile Tensor Accelerator) \cite{VTA-paper,faure2025open}
illustrated in Figure  \ref{fig:VTA} has been designed to accelerate convolutions with ALU and GeMM-based operations. It is a kind of generalization of the TMMA as the user can access VTA instructions and data addresses in the on-chip memory (making it versatile).
 Our formalism can be applied to this architecture as it contains a DRAM, an on-chip memory, and a processing part. However, the strategies, as for the TMMA, must be adapted. The list of applicable accelerators is not exhaustive, but we see that convolution can be modeled as strategies for a large range of accelerators.
\subsection{Outline}

The remainder of this paper is organised as follows:  Section \ref{sec:system} introduces our model for formulating a strategy. Section \ref{sec: convolution} defines the method to decompose a convolution into slices. Building on this formalism, Section \ref{sec: formalisationH1} presents the formalisation of the strategy for S1 and its extended version. Section \ref{sec: OptimH1} then formulates the associated optimisation problem. Section \ref{sec:simu} describes the design and Python implementation of the simulator. Finally, Section \ref{sec: Xp} presents the experimental results and Section \ref{sec-related-work} provides a review of the related work.

\section{System model -- proposed formalism}
\label{sec:system}
Our system model comprises a platform model (CPU, accelerator, on-chip memory, and DRAM) and an application model that details  the computation steps required to execute any application (e.g. convolution layer).

\subsection{Platform model}

The considered accelerator is capable of \nbmac  MAC operations per $t_{acc}$ cycles. The on-chip memory has a size \sizeOnChip. When dealing with memory or data size, we omit the unit (e.g. MB, GB) and only deal with integers that are all expressed in the same unit.
We assume the DRAM is large enough to contain all the data 
required to execute the application (a convolution layer in our case). 
All durations are expressed in the number of accelerator cycles. We assume that the duration to free parts of the on-chip memory is negligible.
We also assume that the duration to load data from the DRAM to the on-chip memory (resp. to write back data from the on-chip memory to the DRAM) is proportional to the size of the data.

We denote the duration to load one element from the DRAM to the on-chip memory by $t_l$ and the duration to write back one element from the on-chip memory to the DRAM by $t_w$.


\subsection{Application model: Steps and actions }
\label{subsec: app model}
The on-chip memory size is rarely sufficient to store all the  data required to compute a convolution layer. As a consequence, the computation is performed in several steps. 

\begin{assumption}
    The on-chip memory is abstracted as a mathematical set, equipped with the set operations: intersection $\bigcap$, union $\bigcup$, set difference$\setminus$ and cardinal $| |$.
    \label{assum:onchipset}
\end{assumption}

\begin{definition}[n-step computation]
\label{def:strategy}
    A n-step computation $S  = (s_1,...,s_n )$ is defined as an ordered execution of steps $s_i$
    \[
    s_i = (F^{inp}_i,  F^{ker}_i, W_i, I^{slice}_i, K^{sub}_i) 
    \]

    where $F^{inp}_i$, $F^{ker}_i$, $W_i$, $I^{slice}_i$, and $K^{sub}_i$ correspond, respectively, to the subset of input part to be freed from the on-chip memory, the subset of kernels to be released from the on-chip memory, the subset of output data to be written back to the DRAM, the subset  input slice loaded from the DRAM to the on-chip memory, and the subset of kernels loaded from the DRAM to the on-chip memory.

\end{definition}

\begin{definition}[Semantics of  a n-step computation]
Each step $s_i=(F^{inp}_i,  F^{ker}_i, W_i, I^{slice}_i, K^{sub}_i)$ is a sequence of actions $\actions_i = (a_1,a_2, a_3, a_4,a_5,a_6 )$ and results in a new on-chip memory state $M_i =
[M_i^{inp},$ $M_i^{ker},M_i^{out}]$ where $M_i^{inp}, M_i^{ker}$ and $M_i^{out}$ correspond, respectively, to the input part, the kernels, and the results that are stored at the end of step $s_i$ in the on-chip memory.
We suppose that the on-chip memory is initially empty, thus
$M_0^{inp}=\emptyset$, $M_0^{ker}=\emptyset$ and $M_0^{out}=\emptyset$.
We use temporary variables
$Mt_i^{inp}$, $Mt_i^{ker}$ and $Mt_i^{out}$ to represent the evolution of the memory.
Practically, a step works as follows:
\begin{enumerate}
    \item 

    $a_1$ removes the input slice $F^{inp}_i$ from on-chip memory, leading to  $Mt_i^{inp} = M_{i-1}^{inp} \setminus F^{inp}_i $;

    \item

    $a_2$ removes parts of the kernels  $F^{ker}_i $
    from on-chip memory, leading to  $Mt_i^{ker} = M_{i-1}^{ker} \setminus F^{ker}_i $;

    \item 

    $a_3$ writes back the computed values $W_i$ in the DRAM, leading to $Mt_i^{out} = M_{i-1}^{out} \setminus W_i$.
    
    \item 

    $a_4$
     an input slice $I^{slice}_i$ is loaded from the DRAM to the on-chip memory.
    Thus, at this moment, the on-chip memory contains the parts of the input
    $M_i^{inp} = Mt_i^{inp} \cup I^{slice}_i$. Indeed, we may keep some input parts in the on-chip memory for further steps and $ Mt_i^{inp}$  represents what was still in the on-chip memory after the action $a_1$;
    
    \item

    $a_5$ a subset of kernels $K^{sub}_i$ is loaded from the DRAM.
     At this moment, the on-chip memory contains the parts of the kernels
    $M_i^{ker} = Mt_i^{ker} \cup K^{sub}_i$. Indeed, we may keep some kernels in the on-chip memory for further steps, and $ Mt_i^{ker}$  represents what was still in the on-chip memory after the action $a_2$
    
    \item 

    $a_6$ the computation is made.
    Once the computation is finished, 
    the result of this computation $Out_i$ 
    is stored in the on-chip memory,
    leading to 
    $M_i^{out} = Mt_i^{out} \cup Out_i$. Indeed, we assume that not all results are written back after a step;
\end{enumerate}
Some sets may be empty (e.g. $I^{slice}_i$, $K^{sub}_i$, ...).
After the very last step $s_n$ the on-chip memory has to be empty and the results have to be written back.
\end{definition}

\begin{definition}[Duration of an n-step strategy]
The duration of one step $s_i$ is a function $f$ of $I^{slice}_i$, $W_i$ and $K^{sub}_i$, i.e. $\durationstep(s_i) = f(I^{slice}_i, W_i,K^{sub}_i)$. Thus, the duration of an n-step strategy is $\displaystyle \durationstep = \displaystyle \sum_{i=1}^{n} \durationstep(s_i)= \sum_{i=1}^{n} f(I^{slice}_i, W_i,K^{sub}_i)$
\label{duration}
\end{definition}

We assume that the duration is proportional to the size of the data involved, we can compute at any time the size (in the expected unit) of the data stored in the on-chip memory using the cardinal function of a set of data (or a memory). Thus :
\begin{itemize}
    \item 
    the duration of a step $s_i$,   $\durationstep(s_i) =$ 
 \[
(|(I^{slice}_i)| + |(K^{sub}_i)|) \times t_l 
     + |(W_i)|\times t_w + t_{acc}
 \]
\item the size of data stored in the on-chip memory during $s_i$:
 \begin{align*}
       size^{step}_i =|M_{i-1}^{inp} \cup I^{slice}_i|  + |M_{i-1}^{ker} \cup K^{sub}_i|\\
       + |M_{i-1}^{out} \cup Out_i|
 \end{align*}
\end{itemize}
 
\subsection{Assumptions}
\label{sec-assumtions-h1}
We consider the same assumptions as the paper \cite{siu-memory-2018}
that are the following:
\begin{itemize}
   
    \item 
    Each input and kernel must be loaded into the on-chip memory from the DRAM a bounded number of times. 
    Indeed, the higher the number, the higher the memory exchange (and possibly the higher the duration). 
    Subsequently, we fix this number to 
    at most twice;

    \item 
    The data loaded in the on-chip memory has to be directly processed by the accelerator. This ensures to not waste memory or in other word, the on-chip memory is sized to contain data that could be computed immediately.

    \item
    The compute action must consume all the inputs and kernels parts stored in the on-chip in a unique execution. Thus, the number of operations to be done in a step must be lower than the number of operations provided by PE, i.e. \nbmac.
  \end{itemize}


\section{Representing the slices for convolutions }
\label{sec: convolution}
The purpose of this section is to remind how convolutions are computed and how they can be decomposed into slices for offloading part of them on the accelerator.

\subsection{Reminder on convolutions}

In general, convolutional layers manipulate tensors in order to represent input and output data, as well as kernels.

\begin{definition}[Tensor]
A tensor $T$ is a multi-dimensional array defined as: $$T \in \mathbb{R}^{n_1 \times n_2\times ...\times n_k}$$
We denote by $T_{i_1,i_2,...,i_k}$ the value of the tensor at indices
$(i_1, i_2,..., i_k) \in [\![0, n_1-1]\!]\times [\![0, n_2-1]\!]\times ...\times [\![0, n_k-1]\!]$.
\end{definition}

\begin{definition}[2D convolution operation]
A 2D convolution operation is a function $f_c$ defined : 
\begin{equation}
\begin{aligned}
f_c : \mathbb{R}^{C_{in} \times H_{in} \times W_{in}} \times \left( \mathbb{R}^{C_{in} \times H_K \times W_K} \right)^N &\to \mathbb{R}^{N \times H_{out} \times W_{out}} \\
(I, \Lambda) &\mapsto O
\end{aligned}
\end{equation}
The function $f_c$ computes the cross-correlation between the input tensor $I$ and the kernel set $\Lambda$ according to the strides and the paddings.
\end{definition}

\begin{remark}
    Note that, although it is called a 2D convolution layer, it takes 3D input because the '2D' refers to the kernel, which moves along two spatial axes on the input ($H_{in}$ and $W_{in}$).
\end{remark}
     \begin{definition}[3D-Input tensor] \
     A \ 3D-input \ tensor
    $I \in $\ $\mathbb{R}^{C_{in} \times H_{in} \times W_{in}}$
    is of dimension $C_{in}$ channels, 
    $H_{in}$ height and $ W_{in}$ width.
     \end{definition}
 A 2D-convolution is parametrized by a set of $N$ kernels $K^i$.
     \begin{definition}[Kernels]
         $\Lambda = \{K^0, K^1,..., K^{N-1}\}$ with $K^i$ 
        $i^{th}$ kernel tensor of a convolution layer : $\forall \space i \in [\![0,N-1]\!],$\\$ \space K^{i} \in \mathbb{R}^{C_{in} \times H_K \times W_K}$, wherein $C_{in}, H_K$ and $ W_K$ represent respectively the channel, the height and the width dimensions of the $i^{th}$ kernel tensor. All the kernels in $\Lambda$ share the same dimensions.
     \end{definition}

A 2D-convolution
     computes a 3D output tensor
    \begin{definition}[3D-Output tensor]
       A 3D-output tensor $O \in \mathbb{R}^{C_{out} \times H_{out} \times W_{out}}$ is of dimension $C_{out}$ channels, $H_{out}$ height and $ W_{out}$ width. 
       Those dimensions depend on the input dimension, the kernels size, the stride $s_w$ along the width dimension, 
       the stride $s_h$ along the height dimensions,
       the padding $p_r , p_l, p_t, p_b$ respectively the right, left, 
       top and bottom. We have: 
       \begin{align*}
                  C_{out} =& N\\
                  H_{out} =& \lfloor \frac{H_{in}-H_K+(p_{t}+p_{b})}{s_{h}} \rfloor + 1\\
                  W_{out} =& \lfloor \frac{W_{in}-W_K+(p_{l}+p_{r})}{s_{w}} \rfloor + 1
       \end{align*}
    \end{definition}

\begin{remark}
    For the sake of clarity, in the remainder of this paper, we assume that the input $I$ is already padded when necessary
    
\end{remark}
The output tensor value is given     
$\forall  \space (l,i,j) \in [\![0,C_{out}-1]\!] \times [\![0,H_{out}-1]\!] \times [\![0,W_{out}-1]\!] $ by: 
\begin{align*}
O_{l,i,j} =  \sum_{c =0}^{{C_{in}-1}}\sum_{h =0}^{H_K-1}\sum_{w =0}^{W_K-1} I_{c,i \times s_h +h, j\times s_w+w}\times K^l_{c,h,w} 
\label{eq:eqconv1}
\end{align*}
This equation shows that the computation of each output value is independent of the others, meaning their computation order does not impact the output result. Consequently, sequencing the execution into independent steps on the accelerator is possible, as long as the slicing is adequate.

For the remainder of this paper, we will write the convolution operation as $\emph{*}$.

\subsection{Slicing of convolution}
In this section, the concept of slicing a tensor is defined, and we consider the particular case of slicing the 3D-Input tensor in a way to compute one output element.

\begin{definition}[Tensor slice]
Let $T$ be a kD-tensor.
A slice $ S \in \mathbb{R}^{l_1,...,l_k}$ of $T$
is a kD-tensor with $a_i, b_i  \in[\![0, n_i-1]\!] $\\$ \text{such that } b_i \geq a_i$, $l_i = (b_i -a_i)$  \\ and $\forall (i_1,..., i_k) \in [\![0,l_1-1]\!] \times ...\times[\![ 0, l_k -1]\!]$, 
$S_{i_1,i_2,...,i_k} =  T_{i_1+a_1,i_2+a_2,...,i_k+a_k}$.    
\end{definition}

The slice of the input tensor involved in computing one output element
is called a \emph{patch}.

\begin{definition}[Patch]
Consider a convolution layer with an input tensor $I \in $ $\mathbb{R}^{C_{in} \times H_{in} \times W_{in}} $. Let us denote $\alpha_{h}^{i}$ and $\alpha_{w}^{j}$ as $\alpha_{h}^{i} = s_h \times i$ and $\alpha_{w}^{j} = s_w \times j$. The patch $P_{i,j}$ involved in the computation of the output elements $O_{l,j,i} $ $\\$ $\text{ with} \ l \in [\![0, C_{out}-1]\!]$ is defined by:
        \begin{align*}
            P_{i,j} = &\{I_{j_1,j_2,j_3}|\space \forall \space (j_1,j_2,j_3) \in \\ &[\![0, C_{in}-1]\!]\times [\![\alpha_{h}^{i}, \alpha_{h}^{i} + H_K ]\!]\times [\![\alpha_{w}^{j}, \alpha_{w}^{j} + W_K]\!]  \}
        \end{align*} 
\end{definition}

\begin{example}[Patches of a convolution layer]
We consider a convolution layer composed of an input tensor $I \in \mathbb{R}^{2 \times5 \times 5}$, a kernel subset $\Lambda = \{K^0, K^1\}$ with $K^i \in \mathbb{R}^{2 \times3 \times 3} $. The strides of the convolution are set as follows  $s_h = s_w =1$.

\begin{figure}[h!bt]
        \centering
        \includegraphics[scale=0.4]{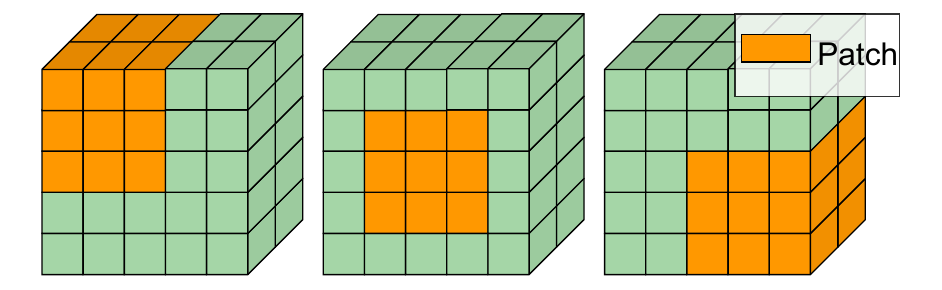} %
        \caption{Example of patches}
        \label{fig:patch}
    \end{figure}

Figure \ref{fig:patch} illustrates three different patches of the input image. The patch on the left corresponds to $P_{0,0}$, the patch in the centre to $P_{1,1}$, and the patch on the right to $P_{2,2}$.
\label{ex:patches}
\end{example}

\begin{definition}[Set of patches]
The set of patches is defined as 
$X = \{P_{0,0}, P_{0,1}, ..., P_{H_{out}-1,W_{out}-1} \}$, X encompasses all the patches necessary to compute a convolution layer.
\end{definition}
\begin{remark}

We consider that $X$ is the set of patch identifiers and
a patch, when manipulated (e.g., to define the slices), is a set of input values.
\end{remark}

\section{Strategy formalization}
\label{sec: formalisationH1}

The work of 
\cite{siu-memory-2018} has proposed four strategies
and we selected one of them, denoted S1-baseline, that offers very interesting settings. 
We encode this baseline with our strategy formalisation and propose an improvement to to minimise the duration \durationstep.

\subsection{S1-baseline formalization}
\label{subsec: form H1}
\begin{figure}[hbt]
        \centering
        \includegraphics[scale=0.250]{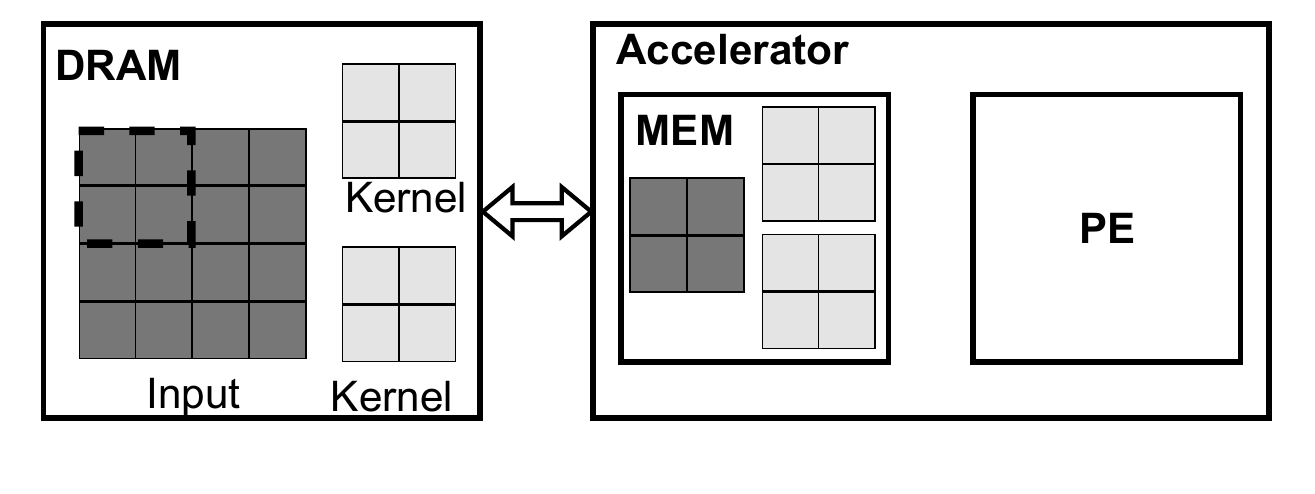} %
        \caption{S1-baseline}
        \label{fig:h1}
    \end{figure}

Figure \ref{fig:h1} shows an example of a step of S1-baseline.
During the first step, all the kernels are loaded into the on-chip memory and they remain there until the last step.

\begin{assumption}\label{assum:patch}
    At each step, one patch of the input is loaded.   
\end{assumption}
The strategy requires $n=|X| = H_{out} \times W_{out}$ steps to compute a convolution.

Let us formalise the strategy S1-baseline with our definition \ref{def:strategy}.
\begin{definition}[n-step computation of S1-baseline from \cite{siu-memory-2018}]
\label{def:heuristic-h1}
Each step consists of computing the convolution of a patch with all kernels. 
A step $s_i = (F^{inp}_i,  F^{ker}_i, W_i, I^{slice}_i,K^{sub}_i)$
is defined as follows.
\begin{itemize}
    \item $I^{slice}_i$, $F^{inp}_i$:
    we have $I_1^{slice} = P_{k_1,l_1}$ where $P_{k_1,l_1}\in X$ is one patch. 
    Let us denote by $X^{done}_i$ the set of patches already computed until step $i$. 
    We have $X^{done}_1=\{P_{k_1,l_1}\}$.
    Then, for $i>1$, we have $I^{slice}_i = P_{k_i,l_i} \setminus M_{i-1}^{inp}$ where $P_{k_i,l_i}\in X\setminus X^{done}_{i-1}$ is a patch not treated. $X^{done}_i=X^{done}_{i-1} \cup \{P_{k_i,l_i}\}$.
    $F_i^{inp} = M_{i-1}^{inp} \setminus P_{k_{i},l_{i}}$.



    \item $K^{sub}_i$, $M^i_{ker}$, $F^{ker}_i$: 
    All the kernels are loaded into the on-chip memory during the first step 
    and never free until the last step. Thus, $K^1_{sub} = \Lambda$ and $K^{i}_{sub}=\emptyset$ for $i>1$.
    We also have $M^i_{ker} = \Lambda$ for $i<n$,
    $F^{ker}_n=  \Lambda$ and  $F^{ker}_i=  \emptyset$ for $i<n$. 

    \item In S1-baseline we do not know when the output elements are stored in the DRAM, and this means we do not know for each step the value of $W_i$. 
    
\end{itemize}
\end{definition}

As reflected in the previous definition, 
the authors of \cite{siu-memory-2018} do not specify
1) the order in which the patches are loaded into the on-chip memory;
2) how and when data are freed
and 3) when the computed values are stored back into the DRAM.

In addition, the choice of loading one patch per step is not optimal  as it does not account for the 
accelerator computational capacity.
Let us compute the number of operations needed to compute one output value.
\begin{definition}[One output value demand]
\label{def:number-mac-per-output}
Let $O_{l,i,j}$ be an output value of the output tensor $O$,
the number of MAC operations required to compute $O_{l,i,j}$ is: 
$$\emph{nb\_op\_value} =  C_{in} \times H_K\times W_K$$
\end{definition}

We need to compute the number of operations performed for one patch when complying with the S1-baseline assumptions.
\begin{property}[Number of operations for S1-baseline]
In the assumptions, all kernels are loaded into the on-chip memory,
thus for a step of S1-baseline,
the output values of all output channels are computed.
The number of operations performed in one S1-baseline step is:
$\emph{nb\_op\_value}  \times C_{out}$.
\end{property}

\subsection{S1 formalization}
We propose to transform  S1-baseline into an optimal strategy S1 that shares the assumptions of S1-baseline and those listed in section \ref{sec-assumtions-h1}.
This entails that we want 1) to compute an ideal order but also 2) to load as many patches as the accelerator can handle (up to \nbmac operations).


  Thus, in S1, the maximal number of patches that need to be loaded into the on-chip memory is 
$$ nb\_patches\_max\_S1 =\big\lfloor \frac{\nbmac}{ nb\_op\_value \times C_{out}}  \big\rfloor$$

\begin{definition}[Minimal number of steps $K_{min}$ for S1]
Let $K_{min}$ be the minimal number of steps required to process the convolution layer. A step can handle at most  \\ \emph{nb\_patches\_max\_S1} patches, which means \\ $K_{min} = \lceil\frac{|X|}{nb\_patches\_max\_S1} \rceil$.
    
\end{definition}

\begin{definition}[Minimal number of steps $K_{max}$ for S1]
Let $K_{max}$ be the maximum number of steps required to process the convolution layer. A step handles at least one patch, which means $K_{max} = |X| $.
    
\end{definition}

Therefore, the number of steps to perform the convolution with S1 is $n \in [\![K_{min},K_{max} ]\!]$.

\begin{figure}[hbt]
        \centering
        \includegraphics[scale=0.60]{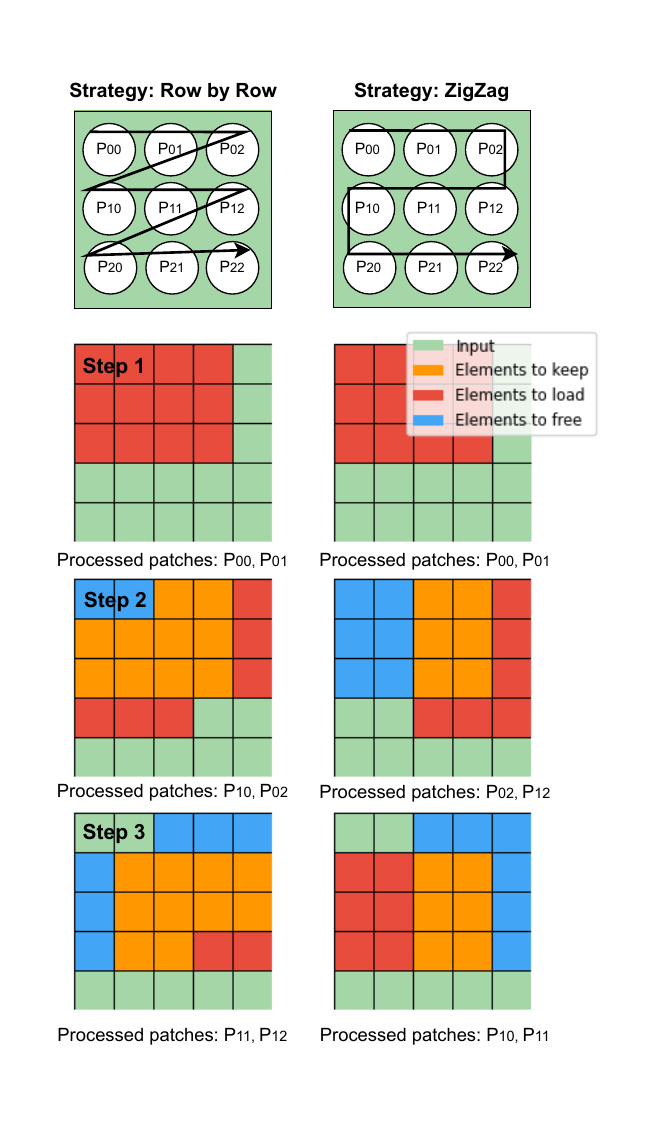} %
        \caption{Example of two different strategies generated by the simulator described in Section \ref{sec:simu}}
        \label{fig:comparaisons-trat}
    \end{figure}

\begin{example}[Row-by-Row VS ZigZag strategies]
 We consider the same convolution layer as in example \ref{ex:patches}. The accelerator is capable of $\nbmac = 120$ MAC operations. Thus, the maximum number of patches that can be processed in a step is $\emph{nb\_patches\_max\_S1} = 2$. Figure \ref{fig:comparaisons-trat} shows two strategies (Row-by-Row and ZigZag) for computing the convolution layer in multiple steps, in these strategies we assume that each output result is written back at the next step.
Let us use our formalism to compare both strategies in the second step: 

\begin{itemize}

\item 

$F^{inp}_2\_Row = \{I_{0,0,0}, I_{1,0,0}, I_{0,0,1}, I_{1,0,1}\}$

$F^{inp}_2\_ZigZag = \{I_{0,0,0}, I_{1,0,0}, I_{0,0,1}, I_{1,0,1}, I_{0,1,0}, I_{1,1,0}, $\\$I_{0,1,1}, I_{1,1,1}, I_{0,2,0}, I_{1,2,0}, I_{0,2,1}, I_{1,2,1} \}$

\item 
$F^{ker}_2\_Row = F^{ker}_2\_ZigZag= \emptyset$

\item 
$W_2\_Row = W_2\_ZigZag =\{O_{0,0,0}, O_{1,0,0}, O_{0,1,0}, $\\$ O_{1,1,0} \}$

\item
    $I^{slice}_2\_Row = \{I_{0,0,4}, I_{1,0,4}, I_{0,1,4}, I_{1,1,4}, I_{0,2,4}, I_{1,2,4}, $\\$ I_{0,3,0},  I_{1,3,0}, I_{0,3,1}, I_{1,3,1}, I_{0,3,2}, I_{1,3,2} \}$

$I^{slice}_2\_ZigZag = \{I_{0,0,4}, I_{1,0,4}, I_{0,1,4}, I_{1,1,4}, I_{0,2,4}, I_{1,2,4}, $\\$I_{0,3,4},  I_{1,3,4}, I_{0,3,3}, I_{1,3,3}, I_{0,3,2}, I_{1,3,2} \}$

\item 
$K^{sub}_2\_Row = K^{sub}_2\_ZigZag =\emptyset$
    
\end{itemize}

Moreover, the memory footprint due to the input for each strategy is: $M_2^{inp}\_Row = 32$ and $M_2^{inp}\_ZigZag = 24$. Finally, we can express the duration of the second step for each step using our duration defined in Section \ref{sec:system}. The size function in this case is the cardinal of a set. Thus, $\durationstep(s_2\_Row) = (|I^{slice}_2\_Row| + |K^{sub}_2\_Row|) \times t_l + |W_2\_Row| \times t_w + t_{acc}  = 6\times t_l + 2 \times t_w+ t_{acc} $ and in the same way \\$\durationstep(s_2\_ZigZag) = 6\times t_l + 2 \times t_w+ t_{acc} $.
\label{example:strat}
\end{example}

Example \ref{example:strat} demonstrates that the order in which patches are computed affects data reuse, traffic between the on-chip memory and DRAM, and the input memory footprint. Therefore, the order of computation is an important factor to consider.

\begin{definition}[n-step strategy of S1]
Each step involves computing a group of patches, $g_i$, where the index i corresponds to the step at which this group is processed. Practically, to determine the group $g_i$, the set is X is partitioned into n partitions, each partition corresponds to a group $g_i$. Thus $G=\{g_0, g_1,...,g_n \} $ where $g_0 = \emptyset$.

A step $s_i = (F^{inp}_i,  F^{ker}_i, W^i, I^{slice}_i,K^{sub}_i)$
is defined as follows:

\begin{itemize}
    \item $I^{slice}_i$, $F^{inp}_i$:
    we have $I_1^{slice} = g_1$ 
    Then, for $i>1$, 
    $$\begin{array}{l}
        I^{slice}_i =  \bigcup_{P_{i,j} \in g_i} P_{i,j} \setminus M^{i-1}_{inp}\\
    \displaystyle F^i_{inp} = M_{i-1}^{inp} \setminus \bigcup_{P_{i,j} \in g_i}P_{i,j}
    \end{array}
   $$


    
    \item $K^{sub}_i$, $M^i_{ker}$, $F^{ker}_i$: 
    All the kernels are loaded into the on-chip memory during the first step 
    and never free until the last step. Thus, $K_1^{sub} = \Lambda$ and $K_{i}^{sub}=\emptyset$ for $i>1$.
    We also have $M_i^{ker} = \Lambda$ for $i<n$,
    $F^{ker}_n=  \Lambda$ and  $F^{ker}_i=  \emptyset$ for $i<n$. 
    
    \end{itemize}

\end{definition}

\section{Optimization problem associated with S1 }
\label{sec: OptimH1}

In this section, we describe how we formulate our optimization problem to find an optimal strategy which respects our constraints and assumptions.
\begin{remark}
The row-major order is used to linearise the indices of the two-dimensional patch into a one-dimensional index. This simplification makes it easier to express the decision variables and constraints in a format that is more suitable for the solver. 
\label{remark:row-major}
\end{remark}
\begin{remark}
The channel-major order is used to linearise the indices of the three-dimensional pixel into a one-dimensional index. This simplification makes it easier to express the decision variables and constraints in a format that is more suitable for the solver. 
\label{remark:channel-major}
\end{remark}

\begin{center}
\begin{tabular}{|l|c|}
    \hline
    \textbf{Variable} & \textbf{Description} \\
    \hline

    \setPixelPatch & Pixel-to-patch assignment \\
    \hline

   \PatchToGroup & Patch-to-group assignment \\
    \hline

   \PixelToGroup & Pixel-to-group assignment  \\
    \hline

    \PixelOverlapped & Presence of a pixel in a group  \\
    \hline

    \PixelIslice & Presence of a pixel in $I^{slice}$  \\
    \hline

    \NbPixelOverlapped & Pixels in commons between $g_i$ et $g_{i+1}$ \\
    \hline
\end{tabular}
\captionof{table}{Variable description} 
\end{center}

\subsection{Problem constants}
In this section, we define the necessary constants to model the optimization problem. The parameters defined in the previous sections, such as the convolution parameters, are also employed throughout this section. To represent the belonging of a pixel to a patch, we define the set : \\ $\setPixelPatch = \{ (i,j) | j \in P_i\}$. This set allows us to model the overlapping between patches and groups. Another constant that we need to define is $nb\_data_\_reload$ which represents the assumption that each input element and kernel element must be loaded into the on-chip memory from the DRAM a bounded number of times.
\begin{remark}\label{remark:2D}
In the definition of $\setPixelPatch$ , we do not take into account the channel dimension. Indeed, to reduce the number of variables, we can work on a 2D tensor when we define the pixels because we do not slice along the channel dimension. 
\end{remark}

\begin{example}
    We consider the same convolutional layer as in Example \ref{ex:patches}. In this example, there are nine patches and 50 pixels. However, as mentioned in Remark \ref{remark:2D}, we do not consider the channel dimension. Therefore, the number of pixels is 25. In this example, therefore, $\setPixelPatch $ is defined as follows:  
    $\setPixelPatch = \{(0,0), (0,1), (0,2), (0,5), (0,6), $ $\\$ $ (0,7), (0,10), (0,11), (0,12),...(8,24) \}$
\end{example}

\subsection{Decision variables with associated constraints}
In this section, we define the necessary decision variables and constraints to model the optimization problem.

\subsubsection{Patches mapping}

The binary decision variable $\PatchToGroup \in \{0,1\}$ represents the presence of a patch $P_{i}$ in the $k^{th}$ group. The variable $\PatchToGroup_{i,k}$ is defined as follows:
\begin{equation}
\PatchToGroup_{i,k} = \begin{cases}
        1 \quad\text{if} \quad P_{i} \in g_k \\
        0 \quad\text{otherwise}
    \end{cases}
\end{equation}

In this strategy, a patch must be assigned to a single group. Therefore, the sum of $\PatchToGroup$  over the number of groups must be equal to one.
\begin{equation}
  \forall i \in [\![0, H_{out} \times W_{out} -1]\!],  \sum_{k=0}^{K_{max}} \PatchToGroup_{i,k} = 1  
  \label{eq:contrainte1}
\end{equation}

Moreover, the cardinality of a group cannot exceed the maximum number of patches defined by \emph{nb\_patches\_max\_S1}. This is because \emph{nb\_patches\_max\_S1} is defined as the maximum number of patches that the accelerator can process in one step, so each group cardinal must be less than or equal to \emph{nb\_patches\_max\_S1}.
\begin{multline}
    \forall  k \in [\![0, K_{max}]\!], \\
      \sum_{i} \PatchToGroup_{i,k} \leq {nb\_patches\_max\_S1}  
  \label{eq:contrainte2}
\end{multline}
\subsubsection{Pixels mapping}

The binary decision variable $\PixelToGroup \in \{0,1\}$ represents the presence of a $pixel_j$ in the $k^{th}$ group. This decision variable is induced by the choice of $\PatchToGroup$. The variable $\PixelToGroup_{
j,k}$ is defined as follows:
\begin{equation}
\PixelToGroup_{j,k} = \begin{cases}
        1 \quad\text{if} \quad pixel_{j} \in g_k \\
        0 \quad\text{otherwise}
    \end{cases}
\end{equation}

The constraint that induces the value of \emph{\PixelToGroup} according to $\PatchToGroup$ is defined as follows:
\begin{equation}
   \PixelToGroup_{j,k} = \bigvee_{i \in |X|} \PatchToGroup_{i,k} 
\end{equation}
This constraint is not linear due to the OR operation, but it can be linearised using well-known techniques \cite{luenberger1984linear}.

\subsubsection{Pixel overlapped}
The binary decision variable $\PixelOverlapped \in \{0,1\}$ represents the presence of a $pixel_j$ in the $k^{th}$ group and $(k-1)^{th}$ group. This decision variable is induced by the choice of $\PatchToGroup$. The variable $\PixelOverlapped_{j,k}$ is expressed as follows:
\begin{multline} 
\PixelOverlapped_{j,k} = \begin{cases}
        \PixelToGroup_{j,k} \wedge  \PixelToGroup_{j,k -1} \quad\text{if} \quad k \geq 1 \\
        0 \quad\text{otherwise}
    \end{cases}
\end{multline}
The AND operator can also be linearised using well-known techniques \cite{luenberger1984linear}.

\subsubsection{Pixel in $I_{slice}$ }
The binary decision variable $\PixelIslice\in \{0,1\}$ represents the presence of a $pixel_j$ in the $k^{th}$ slice of the input ($I_{slice}^k$). This decision variable is induced by the choice of $\PatchToGroup$. The variable $\PixelIslice_{j,k}$ is expressed as follows:
\begin{equation}
 \PixelIslice_{j,k} =
        \PixelToGroup_{j,k} \wedge \neg \PixelOverlapped_{j,k}   
\end{equation}
Using this decision variable, we can express the constraint that each input element and kernel element must be loaded into the on-chip memory from the DRAM a bounded number of times.
\begin{equation}
    \sum_k \PixelIslice_{j,k} \leq nb\_data_\_reload
\end{equation}
\subsection{Decision expressions with associated constraints}
To simplify the expression of our optimisation problem, we introduce decision expressions to make the model more readable and to define useful variables.

\subsubsection{Number of pixels that overlap between two consecutive groups}
The decision expression $\NbPixelOverlapped \in \mathbb{N}$ represents the number of pixels in common between two consecutive groups and is defined as follows:
\begin{equation}
    \NbPixelOverlapped_{k} = \sum_{j=0}^{C_{in}\times H_{in} \times W_{in}-1} \PixelOverlapped_{j,k}
\end{equation}

\subsubsection{Size of a group}
The decision expression $size\_group \in \mathbb{N}$ represents the number of pixels in a group and is defined as follows:
\begin{equation}
    size\_group_{k} = \sum_{j=0}^{C_{in}\times H_{in} \times W_{in}-1} \PixelToGroup_{j,k}
\end{equation}
Now, we can express the constraint that ensures that the data in the on-chip memory fit at any given step. The constraint is defined as follows: 
\begin{multline}
    \forall k \in [\![0, Kmax]\!], size\_group_{k} \\ +  C_{out}\times C_{in} \times H_K \times W_K + \sum_{i} \PatchToGroup_{i,k} \times C_{out} \leq \sizeOnChip
\end{multline}
The first, second and last terms respectively represent the memory footprint due to input, kernels and computed output.

\subsubsection{Size of \texorpdfstring{$I_{slice}^k$}{Islicek}}
The decision expression $size\_I_{slice}^k \in \mathbb{N}$ represents the number of pixels to load in the on-chip memory in step k :
\begin{equation}
  size\_I_{slice}^k = \sum_j \PixelIslice_{j,k}  
\end{equation}

\subsection{Objective function}
Our goal is to minimize the duration of executing an n-step strategy defined in Definition \ref{duration}. Thus, our objective function is as follows:
\begin{equation}
    \durationstep^*=\min_{I^{slice}_i, W_i,K^{sub}_i}(\sum_{i=1}^{n} f(I^{slice}_i, W_i,K^{sub}_i))
\end{equation}

However, we use the duration proportional to the data size defined in Section \ref{sec:system}. Moreover, in this optimisation, we are dealing with S1, meaning that all the kernels are in the on-chip memory. This allows us to assume that the kernels are preloaded in the on-chip memory and that the duration for loading them is not taken into account. Furthermore, we assume that at each step, the computed result is written back into the DRAM, meaning that it is not taken into account in the objective function. Thus, the objective function can be written as follows in this case: 
\begin{multline}
\durationstep^*=\min_{I^{slice}_i}(t_l \times \sum_{i=1}^{n} size(I^{slice}_i) + n\times t_{acc})
\end{multline}

\section{Simulator design and Implementation}
\label{sec:simu}

\begin{figure}[hb!t]
        \centering
        \includegraphics[scale=0.6]{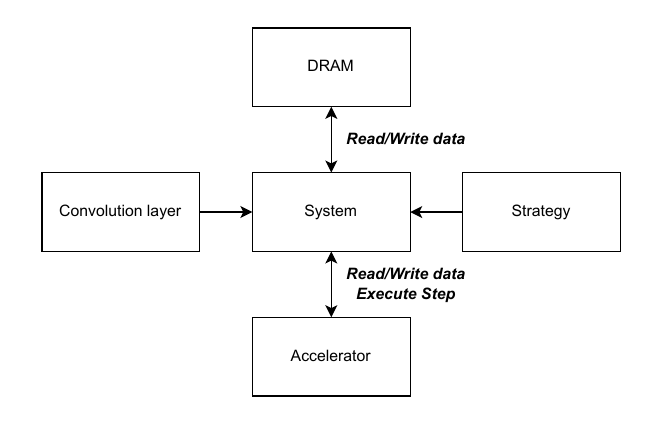} %
        \caption{Diagram of the simulator}
        \label{fig:archi-simu}
    \end{figure}
Based on the formalism of section \ref{sec:system} we have implemented a Python-based simulator that enables a user-defined strategy to be executed on a generic accelerator architecture. This simulator enables users to compare different strategies and visualise the execution of convolutions. Figure \ref{fig:comparaisons-trat} is an example of a visualisation from the simulator. The simulator inputs are: the convolution layer description (stride, layer dimensions, kernel dimensions, number of kernels and the padding) and the data necessary to compute (input tensor and the kernel tensors). It also requires the hardware description (DRAM size, on-chip memory size and accelerator compute capacity) and a strategy that is user-defined or from an ILP solver CSV file. The outputs of the simulator are: a step-by-step execution, a visualisation of the step-by-step strategy, assessment of different metrics (duration and the memory footprint) and a functional simulation that can assess if the result of the step-by-step convolution is correct.
Figure \ref{fig:archi-simu} shows the main building blocks of the simulator. The simulator is implemented in Python and composed of several classes. First, the system acts as orchestrator, providing instructions to the other classes at each step from the strategy instance; 1) reads the current step from the strategy, 2) frees the unnecessary elements in the on-chip memory, 3) writes the results to the DRAM, 4) loads the necessary elements from DRAM to on-chip memory, 5) triggers the accelerator to perform the computation 6) loops to 1). The classes that simulate the hardware are: the accelerator class that implements the on-chip memory and computation part and the DRAM class that implements the off-chip memory. The convolution layer class contains all the parameters and data (patches, pixels and kernels) required for computation. The strategy class implements the sequence of user-defined steps. The simulator will be available open-source on GitHub.

\section{Experimentations}
\label{sec: Xp}
\subsection{Experimental Setup}

We implemented the ILP formulation in the Optimization Programming Language (OPL) and solved it with the IBM CPLEX Optimizer (v22.1) \cite{cplex}. All experiments were run on a 24-core Intel(R) Xeon(R) CPU E5-2420 0 @ 1.90GHz.

\textbf{OPL setup:}
\begin{itemize}
    \item 
    We set the OPL solver timeout to between 0.5 and 5h. 
    
    \item 
    We use the MIP Start functionality to reduce the solving time and guide the solver toward the optimal solution. This allows us to initialize the decision variables with a feasible starting point. Specifically, we inject a solution from either the ZigZag or Row-by-Row strategy, depending on which was best for the given convolution parameters. 
    \item 
    We use the Solution Polishing feature, which switches the solver's strategy from the Branch-and-Cut algorithm to a genetic algorithm, enabling faster convergence towards an optimal solution. As we provide a feasible initial solution, we have configured this switch to occur after 60 seconds of computation.
    \item
    We constrain the search space to a number of groups equal to $K_{min}$ instead of the upper bound $K_{max}$ which leads to a sub-optimal solution. This choice was made to ensure that the problem could be solved in a reasonable amount of time. Using $K_{max}$ would lead to a combinatorial explosion as the number of decision variables $N_{var} = K_{max} \times(3\times(H_{in} \times W_{in}) + H_{out} \times W_{out})$. Therefore, using $K_{min}$ instead of $K_{max} $ drastically reduces the number of groups and reduces the search space.
\end{itemize}

\textbf{Convolution Layers and Hardware Accelerators assessed:}
We evaluated our ILP model on various convolutional layers, adopting the following configuration:
\begin{itemize}
    \item 
    \textbf{Fixed parameters:} The strides were set to $s_h = s_w = 1$ to maximize data reuse opportunities. The number of kernels was set to 1 as this does not affect the optimization of the S1 strategy, with its dimensions set to $H_K =W_K = 3$. 
    \item
    \textbf{Layers dimensions:} We evaluate input sizes $(H_{in}, W_{in}) \in [\![4,12]\!] \times [\![4,12]\!]$ and considering that $H_{in} = W_{in}$.
    \item 
    \textbf{Accelerator computational capacity:} To assess computational impact the parameter $nb\_patches\_max\_S1$ (representing the number of patches that can be loaded into the on-chip memory and directly processed by the accelerator) was varied from 2 to 10. 
    \item 
    \textbf{Memory assumptions:} We assume sufficient the on-chip memory to store all kernels, the $\\$ $nb\_patches\_max\_S1$ patches, and the resulting output for each step.
    
\end{itemize}
\textbf{Metrics:}
The metric used to compare the strategies (ZigZag, Row-by-Row, OPL strategy) is the duration \durationstep. We consider the duration, to be linear and use the definition given in Section \ref{subsec: app model}. As we are evaluating S1 and considering that each output result is written at each step. Thus the duration for performing the computation of the convolutional layer is $\durationstep = t_l \times \sum_{i=1}^{n} size(I^{slice}_i) + n\times t_{acc}$, we set $t_l=t_{acc} =1$. Therefore, the duration is $\durationstep = \sum_{i=1}^{n} size(I^{slice}_i) + n $.



\subsection{ZigZag vs Row-by-Row strategies}
We compare the ZigZag and Row-by-Row strategies on the convolutional layers of ResNet8 and LeNet-5. All experiments were conducted using the simulator. The Row-by-Row strategy operates by grouping $nb\_patches\_max\_S1$ patches sequentially from left to right for every row, as shown in Figure \ref{fig:comparaisons-trat}.
In contrast, the ZigZag strategy organizes the patches by processing even rows from left to right order and odd rows from right to left order, alternating directions, as shown in Figure \ref{fig:comparaisons-trat}.

\begin{figure}[h!]
    \centering
    \includegraphics[width=1.0\linewidth]{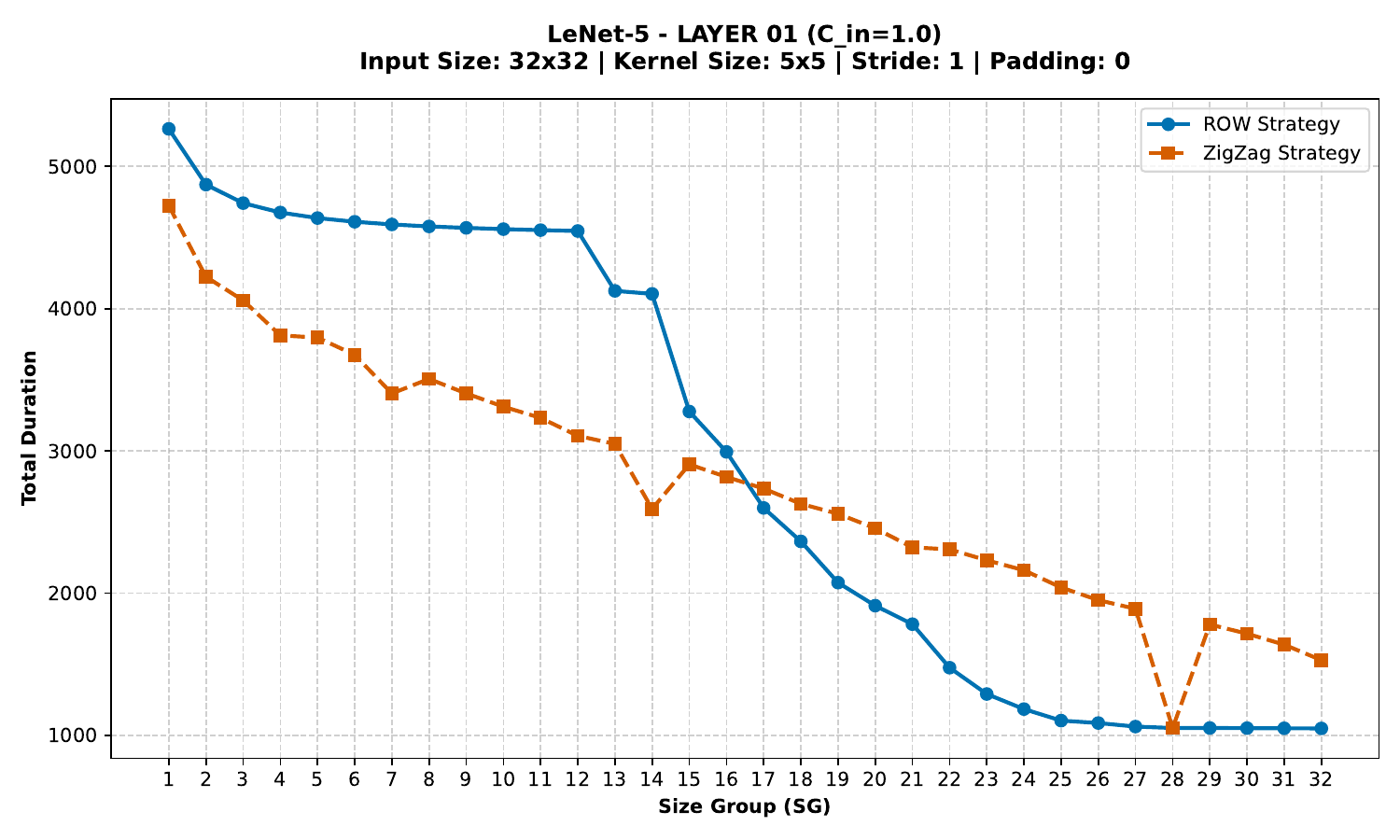}
    \caption{Comparison of the duration of the first LeNet-5 layer using the ZigZag and Row-by-Row strategies for different group sizes.  }
    \label{fig:Latency LeNet5 layer 1}
\end{figure}

The results shown in Figure \ref{fig:Latency LeNet5 layer 1} and other experiments not presented in this work on other convolutional layers demonstrate that the curves have the same shape for ZigZag and Row-by-Row. Secondly, we notice that for small group size, ZigZag outperforms Row-by-Row, whereas after the crossover point the trend reverses. 
A special case happens for group sizes that are a multiple of $W_{out}$ because ZigZag and Row-by-Row strategies are identical. Our simulator rapidly determines the optimal strategy based on the accelerator's constraints.
Furthermore, our experimental results suggest that the crossover point at which Row-by-Row outperforms ZigZag depends on $W_{out}$ and the stride. 

\subsection{ILP Evaluation}

We compare the strategies found by the ILP solver against ZigZag, Row-by-Row and S1-baseline from \cite{siu-memory-2018}. 

\begin{figure}[h!]
    \centering
    \includegraphics[width=1\linewidth]{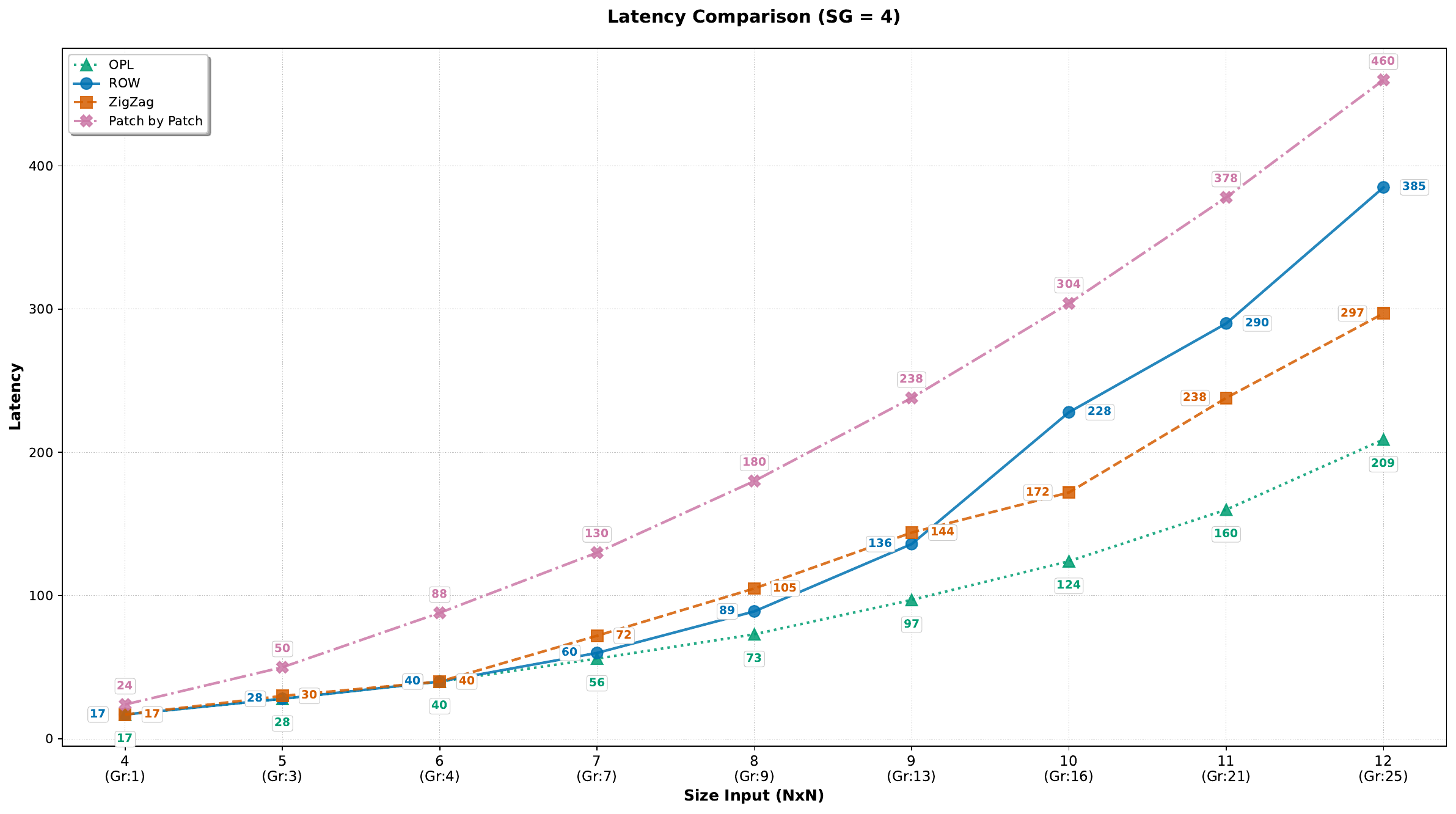}
    \caption{Comparison of the duration for a group size of 4 using the OPL, ZigZag  Row-by-Row and S1 strategies for different input sizes. }
    \label{fig:Latency SG4}
\end{figure}

\begin{figure}[h!]
    \centering
    \includegraphics[width=1\linewidth]{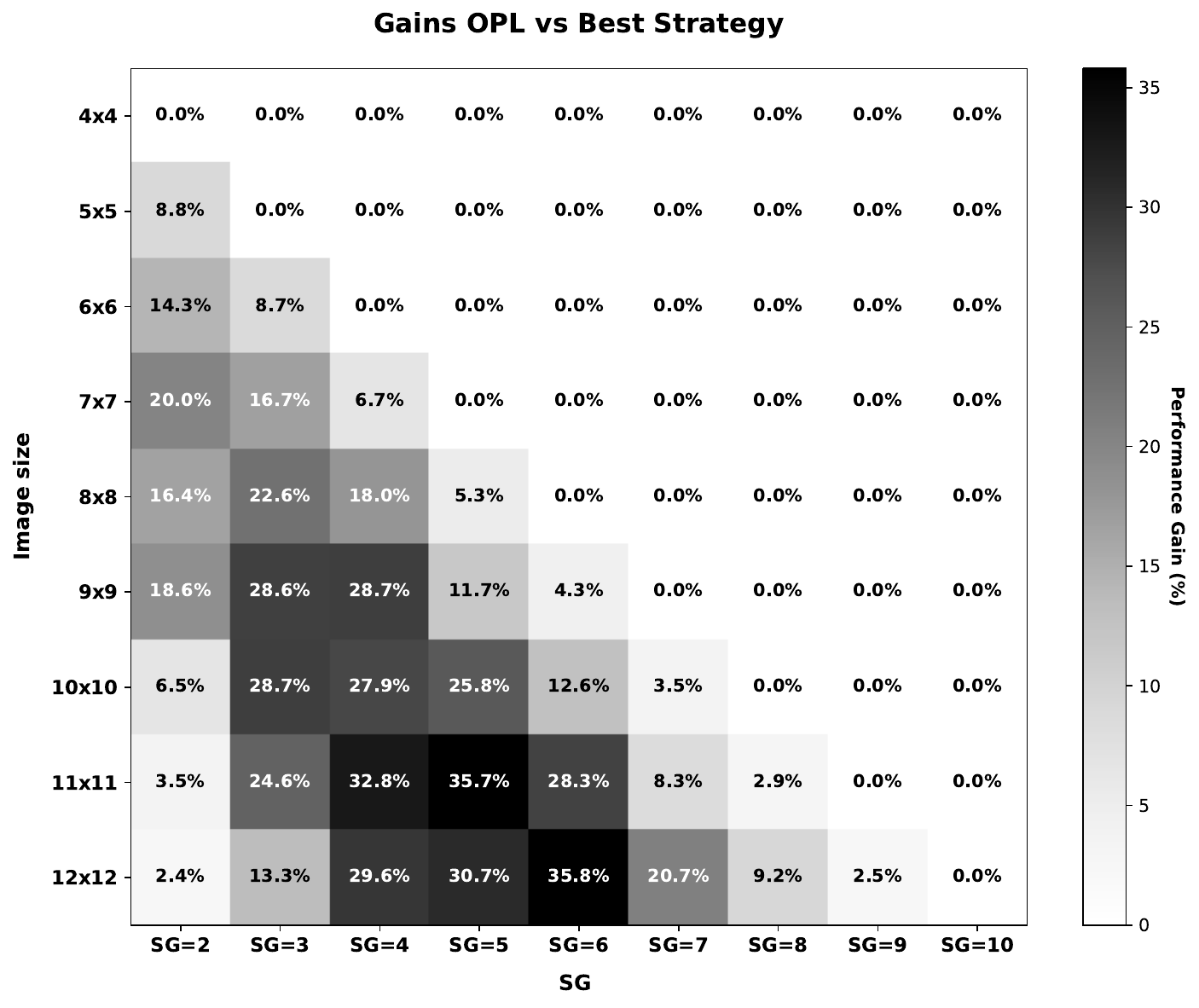}
    \caption{The performance gain between the OPL strategy and the best ZigZag/Row-by-Row strategy.}
    \label{fig:Perf-gain}
\end{figure}

The results shown in Figure~\ref{fig:Latency SG4} and Figure \ref{fig:Perf-gain} demonstrate the efficiency of the ILP formulation. Our method finds a strategy that minimizes the duration $\durationstep$ more efficiently than the other strategies considered.
In Figure \ref{fig:Perf-gain} we observe 2 distinct regions: the upper-right region where is no performance gain (0\% ) corresponds to the case  where the group size (SG) is larger than the image size. Consequently, filling each group becomes trivial. Therefore, ZigZag and Row-by-Row already give an optimal solution. Conversely, the lower-left region highlights cases where the ILP formulation outperforms other strategies, with performance gain reaching up to 30\% .

\section{Related Work}
\label{sec-related-work}



    

In related work, CoSA \cite{CoSA} and LEMON \cite{LEMON} propose to address the problem of scheduling DNN layers on generic specialized accelerators featuring a PE array and a multi-level memory hierarchy. To schedule the computation, the authors use a constrained optimization approach to simultaneously solve optimal loop tiling, loop permutation, and spatial mapping in a single pass. The scheduling space is formulated as a prime-factor allocation problem where the solver determines the optimal allocation of these factors to maximize data reuse and satisfy the memory capacity constraints. Built upon this, LEMON extends the model by considering bandwidths constraints between different memory levels.


At a coarser granularity, COSMA \cite{COSMA} proposes addressing the problem of scheduling DNNs on similar to the generic accelerator architecture we consider but operates at the inter-layer level, considering the layer as the minimal unit, unlike us that optimise within a layer. Their goal is to minimize access to off-chip memory while jointly considering the scheduling of operators, memory allocation, and tensor replacement across the entire DNN graph.
Daini et al.\cite{DainiLZH25} propose scheduling the inference of a CNN model on a multicore system with local SPM and shared DRAM. To achieve this, the authors propose an ILP implementation to optimise the allocation of computational and memory operations on each core, while ensuring that real-time constraints are met. As COSMA, they work at a different granularity level compared to us. Indeed, the authors optimise either layer by layer or the complete CNN model. Unlike us, they take into account the contiguity constraint of loading in the local SPM but our formalism enables to model different architectures compared to them.

In addition to ILP methods for scheduling Deep Neural Networks on specialized accelerators, heuristics and meta-heuristics techniques are employed to explore the vast solution space. GAMMA \cite{GAMMA} and MEDEA \cite{MEDEA} employ genetic algorithms to explore the solution space. Reinforcement learning can also be used for this purpose as demonstrated in \cite{ceiba} \cite{confuciux}. Based on our formalism, we could use these techniques to solve our problem instead of ILP. However, this does not guarantee finding an optimal solution. Another approach to solve this problem is to use gradient-based search methods such as MindMapping \cite{hegde2021mind}. However, this technique is incompatible with our formalism, as it requires the problem to be differentiable, whereas the current problem formulation is mainly discrete.

 The most ubiquitous technique to compute convolution is General Matrix Multiplication (GeMM). This technique relies on efficient matrix multiplication (e.g. the Strassen algorithm). The input tensor is transformed into a matrix using the im2col algorithm. A drawback of this technique is that it involves memory overhead as the overlapping elements between patches are duplicated in the input matrix. Moreover, each patch defined in Section \ref{sec: convolution} corresponds to a distinct column of the input matrix; therefore, our formalism can be adapted to this technique. However, the sequence of steps found by the ILP solver cannot be used, as the input elements are duplicated, meaning there is no opportunity for data reuse. For future work, we plan to adapt both our formalism and our ILP formulation for the GeMM algorithm.

Finally, there exist other techniques for computing convolution other than the classical one \cite{dealbuquerquesilva} \cite{jorda2019performance}. The Fast Fourier Transform (FFT)-based convolution leverages the Fourier transform to project both the input and kernel tensors into the frequency domain.  This reduces the number of required operations since convolution in the frequency domain becomes a Hadamard product. However, FFT-based convolution requires the input and kernel tensors to have the same dimensions, which necessitates zero-padding the kernel. Therefore, FFT-based convolution is only computationally more efficient than other techniques if the kernel size is large, which rarely occurs in state-of-the-art CNNs. In fact, CNNs mainly rely on kernel sizes of 1x1 and 3x3.
The second is Winograd-based convolution. Similarly to FFT-based convolution, it projects the tensors into a polynomial basis. Winograd-based convolution is efficient for small kernels (3x3 and 1x1), but becomes unstable numerically as the size of the input and kernel increases. 
\section{Conclusion}\label{sec:conclusion}


We proposed a formalism to define a strategy of computation on a generic accelerator. We applied this formalism to a strategy from the literature. We then extended it to take into account the computation capacity of the accelerator. Based on the formalism, we formulated an ILP problem to minimise the duration $\durationstep$ for computing a convolution while meeting the on-chip memory size constraint. Our results demonstrate that the ILP solution is more efficient than the ZigZag and Row-by-Row strategies. Finally, we developed a Python-based simulator to analyse strategies. This has enabled us to compare the ZigZag and Row-by-Row strategies and to assess the solutions found by the ILP solver.

In future work, we plan to present the formalism and ILP formulation of other offloading strategies that operate at a finer granularity than patches and do not assume that all kernels are stored in on-chip memory during computation. Secondly, in our current model we have completely abstracted the memory organization (assumption \ref{assum:onchipset}). To offload a part of the input and have a linear duration, this supposes that the data are packed into the correct order. We plan to provide a more representative memory model similar to the one in \cite{DainiLZH25}. Finally, to address larger model, we plan to find a hybrid solution combining heuristics and ILP.



\section*{Acknowledgment}
This work has benefited from the AI cluster ANITI2 funded by the French government through the ANR under the France 2030 program (grant ANR-23-IACL-0002).

\bibliographystyle{abbrv}
\bibliography{bib}

\end{document}